\documentclass[aps,prd,twocolumn,groupedaddress,showpacs,nofootinbib,amssymb]{revtex4}
\usepackage{graphicx,bm,color}
\usepackage{amsmath}
\usepackage{amssymb}
\usepackage{amsfonts}
\usepackage{cases}
\usepackage{color}
\allowdisplaybreaks[4]

\begin{document}
\title{Stability and Anti-evaporation of the Schwarzschild-de Sitter Black Holes in Bigravity }
\author{Taishi Katsuragawa$^{1}$ and Shin'ichi Nojiri$^{1,2}$ }
\affiliation{
$^1$ Department of Physics, Nagoya University, Nagoya
464-8602, Japan \\
$^2$ Kobayashi-Maskawa Institute for the Origin of Particles and
the Universe, Nagoya University, Nagoya 464-8602, Japan 
}

\begin{abstract}
We study the stability under the perturbation and the related anti-evaporation 
of the Nariai space-time in bigravity.
If we impose specific condition for the solutions and parameters, 
we obtain asymptotically de Sitter space-time, 
and show the existence of the Nariai space-time as a background solution.
Considering the perturbation around the Nariai space-time up to first order,
we investigate the behavior of black hole horizon.
We show that the anti-evaporation does not occur on the classical level in the bigravity.
\end{abstract}

\pacs{04.62.+v, 98.80.Cq}

\maketitle

\section{Introduction}

One can find that there exist many reasons and motivations to consider alternative theories 
of gravity to the general relativity.
Some theories are motivated by the modifications in infrared regime and 
they mainly aim to resolve a question about the dark energy.
For instance, $F(R)$ gravity \cite{Nojiri:2003ft,Nojiri:2006gh} can explain the accelerating 
expansion of current universe without
the cosmological constant and avoid the hierarchy problem.
Others are motivated by the modifications in ultraviolet regime and 
they are often associated with the effects of quantum gravity 
\cite{'tHooft:1974bx,Donoghue:1994dn}.
Higher-curvature theories typified by curvature-squared and the Gauss-Bonnet terms
are induced from quantum corrections.
Naturally, challenges to the theory beyond general relativity themselves are important 
because there is no fundamental reason to choose the Einstein-Hilbert action 
or Einstein's equation
over many kind of alternatives.

Recently, much attention has been paid to bi-metric theory or what we call bigravity, 
which includes two independent metric tensor fields, $g_{\mu \nu}$ and $f_{\mu \nu}$
\cite{Hassan:2011hr,Hassan:2011vm,Hassan:2011zd}.
Bigravity contains massive spin-$2$ propagating mode 
in addition to ordinary massless spin-$2$ mode corresponding to the graviton.
This theory has been successfully constructed as the generalization of dRGT 
massive gravity in recent years
\cite{deRham:2010ik,deRham:2010kj}.
Some people expect that the new degrees of freedom introduced by another metric can solve
remaining problems in cosmology, that is, dark energy 
\cite{Damour:2002wu,Volkov:2011an,Volkov:2012zb,vonStrauss:2011mq,Akrami:2012,Berg:2012kn,
Volkov:2013roa,Akrami:2013ffa,Tamanini:2013xia,Fasiello:2013woa,
Akrami:2014ff,Konnig:2014fj}
and dark matter 
\cite{Rossi:2009ju,Banados:2008fi,Aoki:2013joa,Aoki:2014cla}
problems.
Interactions between two metric tensors produce effective cosmological constant, furthermore,
the massive spin-$2$ fields and matter fields coupled with the metric $f_{\mu \nu}$ 
can be candidates of dark matter.

When we intend to view the bigravity to be an alternative theory of gravity,
it is also interesting that we apply this theory to other phenomena in cosmology or astrophysics 
and find the differences from general relativity.
In our work, we focus on the nature of black holes.
It is well known that the horizon radius of the black hole 
in the vacuum usually decreases by the Hawking radiation. 
However, Bousso and Hawking have observed a phenomenon where the black hole 
radius increases by the quantum correction for the specific Nariai black hole
\cite{Bousso:1997wi}. 
This phenomenon is called anti-evaporation of black holes. 
Note that the Schwarzschild-de Sitter black holes and the Nariai black holes can be primordial 
ones, thus, they are not expected to appear at the final stage of star collapse.  

%%%%%%%%%%%%%%%%%%%%%%%%%%%%
In order that the anti-evaporation could occur in the Einstein gravity, we need to include the 
quantum correction from the matters. 
It might be remarkable, however, that the anti-evaporation may occur even on the classical level 
in $F(R)$ gravity theories \cite{Nojiri:2013su,Sebastiani:2013fsa,Nojiri:2014jqa}. 
This could be because that the equations for the $F(R)$ gravity are more complicated than those 
for the Einstein gravity, but at present, it is not so clear what could be essential for the 
anti-evaporation on the classical level. 
Then it might be interesting if the anti-evaporation in the classical level might be a general 
phenomena in the modified gravity. 
In this paper, we consider the possibility of the anti-evaporation in bigravity on the classical level
because the contribution from the interaction between two metric tensors is not so trivial.
We will give a classical analysis in the stability of the Nariai black hole in the bigravity
and study if the anti-evaporation could occur in the classical level, which 
may clarify what could be necessary for the anti-evaporation to occur. 
%%%%%%%%%%%%%%%%%%%%%%%%%%%%%%%

This paper is organized as follows: 
%The contents of this paper are following.
First, we explain about the Nariai black hole.
The Nariai black hole is defined as a subset of the Schwarzschild-de Sitter black hole where 
the radii of the cosmological and black hole horizons are degenerate.
Second, we give a brief review about the bigravity. 
In order to show the existence of the Nariai black hole as an exact solution in the bigravity, we 
specify some parameters and solutions, 
then, we give a proof that the asymptotically de Sitter solutions can be realized.
Finally, we consider the perturbations around the black hole and evaluate their stability 
and investigate if the anti-evaporation could occur on the classical level. 

\section{Anti-evaporation of the Nariai Black Holes}

\subsection{Nariai space-time and its property}

At first, we introduce the Nariai space-time as a family of the Schwarzschild-de Sitter space-time.
The Schwarzschild-de Sitter solution is expressed in the following form:
\begin{align}
ds^{2} = -V(r)dt^{2} + V(r)^{-1}dr^{2} + r^{2}d\Omega^{2}\, , 
\label{polar coordinate} ,
\end{align}
where the function $V(r)$ is defined by
\begin{align}
V(r) = 1 - \frac{2\mu}{r} - \frac{\Lambda}{3}r^{2}\, .
\end{align}
Here, $\mu$ is a mass parameter and $\Lambda$ is a positive cosmological constant.
For $0 < \mu < \frac{1}{3} \Lambda^{-1/2}$, $V(r)$ has two positive roots $r_{c}$ and $r_{b}$, 
corresponding to the cosmological and black hole horizon, respectively.
In the limit $\mu \rightarrow \frac{1}{3}\Lambda^{-1/2}$, 
the radius of the black hole horizon coincide with that of the cosmological horizon.
Here, the coordinate system in Eq.~(\ref{polar coordinate}) becomes inappropriate 
because $V(r) \rightarrow 0$ between the two horizons.
Then it is useful to introduce new coordinate system as follows:
\begin{align}
t = \frac{1}{\epsilon \sqrt{\Lambda}}\psi\, , \quad 
r = \frac{1}{\sqrt{\Lambda}} \left( 1 - \epsilon \cos \chi - \frac{1}{6} \epsilon^{2} \right)\, , 
\end{align}
where $\epsilon$ is the parameter defined as $9\mu^{2} \Lambda = 1 - 3\epsilon^{2}$, 
and $\epsilon \rightarrow 0$ corresponds to the degeneracy of two horizons.

In above coordinate, the black hole horizon corresponds to $\chi = 0$ and the cosmological horizon 
corresponds to $\chi = \pi$, and the metric takes the following form:
\begin{align}
ds^{2} =& - \frac{1}{\Lambda} \left( 1 + \frac{2}{3} \epsilon \cos \chi \right) \sin^{2} \chi d\psi^{2} 
\nonumber \\
& + \frac{1}{\Lambda} \left( 1 - \frac{2}{3}\epsilon \cos \chi \right) d\chi^{2} 
+ \frac{1}{\Lambda} ( 1 - 2\epsilon \cos \chi ) d\Omega^{2} \, .
\end{align}
In the degenerate case, $\epsilon = 0$, the metric is given by
\begin{align}
ds^{2} = \frac{1}{\Lambda} \left( - \sin^{2} \chi d\psi^{2} + d\chi^{2} \right) 
+ \frac{1}{\Lambda} d\Omega^{2}\, ,
\end{align}
and this space-time is called the Nariai black hole.
Note that the topology of the space-like sections of the Schwarzschild-de Sitter space-time 
(and the Nariai space-time) is $S^{1} \times S^{2}$ 
while that of the ordinary black hole solution is $S^{2}$ in four dimensions.
In this coordinate system, the radius of two-sphere, $r$, varies along the one-sphere coordinate, 
$\chi$; 
the minimal two-sphere corresponds to the black hole horizon 
and the maximal one corresponds to the cosmological horizon.

\subsection{Trace anomaly and anti-evaporation}

In this section, we give a brief review of the anti-evaporation in general relativity.
First of all, we begin with the Hawking radiation from the black holes.
It is well known that there is radiation by the quantum effects of matter fields
around the black hole horizon, which is called the Hawking radiation.
This quantum corrections leads to the trace anomaly of the energy-momentum tensor
although the trace of the energy-momentum tensor should classically vanish, $T^{\mu}_{\ \mu}=0$.
When we consider the massless scalar as the Hawking radiation, 
the effective action corresponding to the trace anomaly
is written by a covariant form \cite{Bousso:1997cg,Nojiri:1997hx},
\begin{align}
S_{\mathrm{eff}} =& - \frac{1}{48 \pi G}\int d^{2}x \sqrt{-g} \nonumber \\
& \hphantom{- \frac{1}{48 \pi G}\int }
\times \left [ \frac{1}{2}R \frac{1}{\Box} - 6 (\nabla \phi)^{2} \frac{1}{\Box} R - \omega \phi R 
\right ]\, .
\end{align}
Here, the effective action is reduced to two dimensional form,
and $\omega$ is the redundancy parameter corresponding to the %cut-off 
renormalization scheme.

The above effective action leads to the modification for the equation of motion.
In general relativity, 
specific perturbations around the Nariai black holes shrink from its initial values. 
Then, the size of black hole horizon increases at least initially.
This phenomena is called anti-evaporation.
Note that if we do not include the quantum correction, the anti-evaporation does not occur and
the horizon size remains in that of the initial perturbation.

%%%%%%%%%%%%%%%%%%%%%%%%%%%%%%%%%%%%
%In $F(R)$ gravity, equation of motion is modified as different way from general relativity,
%however, anti-evaporation is realized on the classical level, that is, without quantum corrections.
%Our case, bigravity, also modify the equation of motion compared with genera relativity,
%therefore, it is interesting to study if the anti-evaporation occurs in the classical level.
%%%%%%%%%%%%%%%%%%%%%%%%%%%%%%%%%%%%%

We should note that in case of the $F(R)$ gravity , there occurs the anti-evaporation even on 
the classical level, that is, without quantum correction \cite{Nojiri:2013su,Sebastiani:2013fsa}. 
In the $F(R)$ gravity, the equations are complicated, which may generate the the anti-evaporation 
on the classical level. 
Because the equations in the bigravity are also pretty complicated, 
we may expect that the anti-evaporation could occur on the classical level, and therefore, 
it could be interesting to investigate the anti-evaporation in the bigravity even if on the classical level. 

%%%%%%%%%%%%%%%%%%%%%%%%%%%%%

\section{Nariai Black Holes in Bigravity}

In this section, we give a brief review of the bigravity and 
show that the Nariai space-time is an exact solution in this theory.
The action of the bigravity \cite{deRham:2010kj} is given by
\begin{align} 
S_\mathrm{bigravity}  
& = M^{2}_{g} \int d^{4}x \sqrt{-\mathrm{det}(g)}R(g) \nonumber \\
& + M^{2}_{f} \int d^{4}x \sqrt{-\mathrm{det}(f)}R(f)  \nonumber \\
& - 2m^{2}_{0} \, M^{2}_\mathrm{eff} \int d^{4}x \sqrt{-\mathrm{det}(g)} 
\sum^{4}_{n=0} \beta_{n}e_{n} \left( \sqrt{g^{-1}f} \right)  \, .
\label{bigravity action}
\end{align}
Here, $g$ and $f$ are dynamical variables and rank-two tensor fields which have properties as 
metrics,
$R(g)$ and $R(f)$ are the Ricci scalars for $g_{\mu \nu}$ and $f_{\mu \nu}$, respectively,
$M_{g}$ and $M_{f}$ are the two Planck mass scales for $g_{\mu \nu}$ and $f_{\mu \nu}$ as well, 
and the scale $M_\mathrm{eff}$ is the effective Planck mass scale defined by
\begin{align}
\frac{1}{M^{2}_\mathrm{eff}} = \frac{1}{M^{2}_{g}} + \frac{1}{M^{2}_{f}}\, .
\label{Meff}
\end{align}
The quantities $\beta_{n}$s and $m_{0}$ are free parameters, 
and the former defines the form of interactions and the latter expresses the mass of the massive 
spin-$2$ field.
The matrix $\sqrt{g^{-1}f}$ is defined by the square root of $g^{\mu \rho}f_{\rho \nu}$, that is,
\begin{align}
\left( \sqrt{g^{-1}f} \right)^{\mu}_{\ \rho} \left( \sqrt{g^{-1}f} \right)^{\rho}_{\ \nu} = g^{\mu \rho}f_{\rho \nu}\, . \label{sqrtfg}
\end{align}
For general matrix $\mathbf{X}$, $e_{n}(\mathbf{X})$s are polynomials of the eigenvalues of  $X$:
\begin{align}
e_{0}(\mathbf{X}) =& 1\, , \quad 
e_{1}(\mathbf{X}) = [\mathbf{X}]\,  , \nonumber \\
e_{2}(\mathbf{X}) =& \frac{1}{2} \left( [\mathbf{X}]^{2} - [\mathbf{X}^{2}] \right)\, , \nonumber \\
e_{3}(\mathbf{X}) =& \frac{1}{6} 
\left( [\mathbf{X}]^{3} - 3[\mathbf{X}][\mathbf{X}^{2}] + 2[\mathbf{X}^{3}] \right)\, , \nonumber \\
e_{4}(\mathbf{X}) =& \frac{1}{24} 
\left( [\mathbf{X}]^{4} - 6[\mathbf{X}]^{2}[\mathbf{X}^{2}] + 3[\mathbf{X}^{2}]^{2} \right. \nonumber \\
& \left. \qquad + 8[\mathbf{X}][\mathbf{X}^{3}] - 6[\mathbf{X}^{4}]  \right)  \nonumber \\
=& \mathrm{det}(\mathbf{X})\, , \nonumber \\ 
e_{k}(\mathbf{X}) =& 0 \quad \mbox{for} \ \ k>4 \, , \label{e_n}
\end{align}
where the square brackets denote the traces of the matrices, that is, $[X]=X^{\mu}_{\mu}$.
For conventional notation, we explicitly denote the determinant of matrix $A$ as $\mathrm{det}(A)$, 
and $\sqrt{A}$ represents a matrix which is the square root of $A$.

Now we consider the variation of the action (\ref{bigravity action}) with respect to $g_{\mu \nu}$.
The equation of motion for $g_{\mu \nu}$ is given by
\begin{align}
0 =& R_{\mu \nu}(g) - \frac{1}{2}R(g)g_{\mu \nu} \nonumber \\
&+ \frac{1}{2} \left( \frac{m_{0}M_\mathrm{eff}}{M_{g}} \right)^{2}
\sum^{3}_{n=0}(-1)^{n}\beta_{n} \nonumber \\
& \times \left \{ g_{\mu \lambda}Y^{\lambda}_{(n) \nu}(\sqrt{g^{-1}f}) 
+ g_{\nu \lambda}Y^{\lambda}_{(n) \mu}(\sqrt{g^{-1}f}) \right \}\, . \label{geq} 
\end{align}
Here, for a matrix $\mathbf{X}$, $Y_{n}(\mathbf{X})$s are defined by
\begin{align}
Y^{\lambda}_{(n) \nu}(\mathbf{X})=\sum^{n}_{r=0} (-1)^{r} \left( X^{n-r} \right)^{\lambda}_{\ \nu} 
e_{r}(\mathbf{X})\, ,
\end{align}
or explicitly, 
\begin{align}
Y_{0}(\mathbf{X}) =& \mathbf{1}\, , \quad 
Y_{1}(\mathbf{X}) = \mathbf{X}- \mathbf{1}[\mathbf{X}] \, , \nonumber \\
Y_{2}(\mathbf{X}) =& \mathbf{X}^{2} - \mathbf{X}[\mathbf{X}] 
+ \frac{1}{2} \mathbf{1} \left( [\mathbf{X}]^{2} - [\mathbf{X}^{2}] \right) \, , \nonumber \\
Y_{3}(\mathbf{X}) =& \mathbf{X}^{3} - \mathbf{X}^{2}[\mathbf{X}] 
+ \frac{1}{2} \mathbf{X} \left( [\mathbf{X}]^{2} - [\mathbf{X}^{2}] \right) \nonumber \\
&- \frac{1}{6} \mathbf{1} \left( [\mathbf{X}]^{3} - 3[\mathbf{X}][\mathbf{X}^{2}] 
+ 2[\mathbf{X}^{3}] \right)\, .  \label{interaction_terms}
\end{align}
We also obtain the equation of motion for $f_{\mu \nu}$,
\begin{align}
0 =& R_{\mu \nu}(f) - \frac{1}{2}R(f)f_{\mu \nu} \nonumber \\
&+ \frac{1}{2} \left( \frac{m_{0}M_\mathrm{eff}}{M_{f}} \right)^{2} 
\sum^{3}_{n=0}(-1)^{n}\beta_{4-n} \nonumber \\
& \times \left \{ f_{\mu \lambda}Y^{\lambda}_{(n) \nu}(\sqrt{f^{-1}g}) 
+ f_{\nu \lambda}Y^{\lambda}_{(n) \mu}(\sqrt{f^{-1}g}) \right \} . \label{feq}
\end{align}
In this case, we do not consider the energy-momentum tensor for the ordinary matter fields.
The constraints for the conservation law appear if the minimal couplings to the matter are introduced,
and we find
\begin{align}
0=&\nabla^{\mu}_{(g)} \left[ \sum^{3}_{n=0}(-1)^{n}\beta_{n} 
\left \{ g_{\mu \lambda}Y^{\lambda}_{(n) \nu}(\sqrt{g^{-1}f}) \right. \right. \nonumber \\
& \hphantom{
\nabla^{\mu}_{(g)} [ \sum^{3}_{n=0}(-1)^{n}\beta_{n} \{ g_{\mu \lambda} }
\left. \left. 
+ g_{\nu \lambda}Y^{\lambda}_{(n) \mu}(\sqrt{g^{-1}f}) \right \} \right] \, , \\
0=&\nabla^{\mu}_{(f)} \left[ \sum^{3}_{n=0}(-1)^{n}\beta_{4-n} 
\left \{ f_{\mu \lambda}Y^{\lambda}_{(n) \nu}(\sqrt{f^{-1}g}) \right. \right. \nonumber \\
& \hphantom{
\nabla^{\mu}_{(f)} [ \sum^{3}_{n=0}(-1)^{n}\beta_{4-n} \{ f }
\left. \left.
+ f_{\nu \lambda}Y^{\lambda}_{(n) \mu}(\sqrt{f^{-1}g}) \right \} \right] \, .
\end{align}
Here, $\nabla_{(g)}$ and $\nabla_{(f)}$ are covariant derivatives which are defined in terms of 
$g_{\mu \nu}$ and $f_{\mu \nu}$, respectively. 

In order to discuss the anti-evaporation in the bigravity, 
we need to confirm that the asymptotically de-Sitter solutions are realized in this theory.
However, it is not so easy to investigate this problem for all combinations of the included parameters,
thus, we impose specific assumptions to make discussion simpler.
One of the authors considered a particular class of solutions 
where the two metric tensors are proportional to each other \cite{Katsuragawa:2013lfa} ,
\begin{align}
f_{\mu \nu} = C^{2}g_{\mu \nu}\, . \label{proportional relation}
\end{align}
This proportional relation leads to the Einstein's equation with a cosmological constant 
because $\sqrt{g^{-1}f}$ and $\sqrt{f^{-1}g}$ turn to be proportional to unity.
The two equations of motion are given by
\begin{align}
0 =& R_{\mu \nu}(g) - \frac{1}{2}R(g)g_{\mu \nu} + \Lambda_{g}(C)g_{\mu \nu}  \label{geq1}\, , \\
0 =& R_{\mu \nu}(f) - \frac{1}{2}R(f)f_{\mu \nu} + \Lambda_{f}(C) f_{\mu \nu} \label{feq1}\, .
\end{align}
Note that the dynamics of two metric tensors $g_{\mu \nu}$ and $f_{\mu \nu}$ 
are separated from each other 
and the constraints derived from the preservation of energy-momentum tensor 
are automatically satisfied.
And, we have not assumed any symmetries in the space-time, thus,
we can impose the spherical symmetry to the solutions later.

\begin{figure}
\centering
\includegraphics[width=0.7\columnwidth]{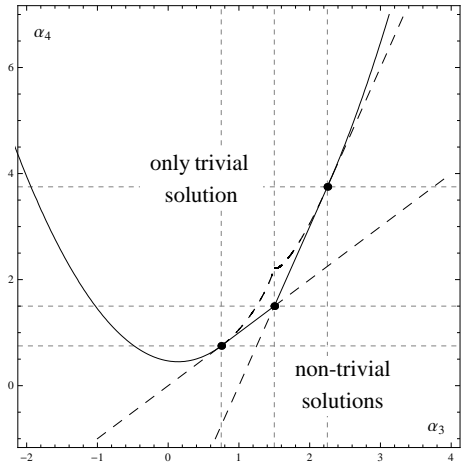}
\caption{Classification of the parameters $\alpha_{3}$ and $\alpha_{4}$ are 
given \cite{Katsuragawa:2013lfa}.
We obtain only flat solutions in the region ``only trivial solution'', 
although asymptotically non-flat (de Sitter and/or anti-de Sitter) solutions are realized in 
``non-trivial solutions''.
}
\label{fig}
\end{figure}

Furthermore, we consider a specific parametrization for the interacting parameter $\beta_{n}$s, 
\begin{align}
&\beta_{0} = 6 - 4 \alpha_{3} + \alpha_{4}\, , \quad
\beta_{1} = -3 + 3\alpha_{3} - \alpha_{4} \, , \nonumber \\
&\beta_{2} = 1- 2\alpha_{3} + \alpha_{4} \, , \quad
\beta_{3} = \alpha_{3} - \alpha_{4} \, , \quad
\beta_{4} = \alpha_{4}\, .
\end{align}
This combination of two parameters, $\alpha_{3}$ and $\alpha_{4}$, are required 
by the existence of the solution corresponding to the flat space-time in massive gravity,
which is often used in bigravity.
With a assumption $M_{g} = M_{f}$, 
it has been shown that the de Sitter solution can be realized in some parameter regions (Fig.\ref{fig}).
For instance, the minimal model $(\alpha_{3}, \alpha_{4}) = (1, 1)$ has only asymptotically flat solutions
although the next-to minimal models $(\alpha_{3}, \alpha_{4}) = (1, -1), (-1, 1), (-1, -1)$ have 
asymptotically de Sitter solutions.
Therefore, the Schwarzschild-de Sitter black hole solutions are realized in the bigravity,
and we can obtain the Nariai black hole solution 
by the limit $\mu \rightarrow \frac{1}{3}\Lambda^{-1/2}$.
Note that it is recently implied that the de Sitter solution is an attractor with homothetic relation,
which may support us to consider our setting.

Under the assumption that $f_{\mu \nu} = C^{2}g_{\mu \nu}$, 
the dynamics of two metric tensors $g_{\mu \nu}$ and $f_{\mu \nu}$ 
are separated from each other, described by Einstein's equations.
Here, this property is just for the background solution
and perturbations can be independent; 
degrees of freedom of bigravity does not descend to that of general relativity.
When we consider the perturbations from a background space-time,
the interaction terms give non-trivial contribution to the evolution 
compared with the case of general relativity.
Therefore, it is important to analyze the stability of perturbation even on the classical level,
and we need to investigate if the anti-evaporation could be realized on the classical level.

\section{Stability of the Schwarzschild-de Sitter black hole}

\subsection{Background solution}

We now consider the perturbation from the Nariai black hole.
According to the topology, $S^{1} \times S^{2}$, we make spherically symmetric metric ansatz as 
follows,
\begin{align}
g_{\mu \nu} dx^{\mu} dx^{\nu}
=& e^{2\rho_{1}(t,x)} \left( -dt^{2} + dx^{2} \right) \nonumber \\
&+ e^{-2\varphi_{1}(t,x)}\left( d\theta^{2} + \sin^{2} \theta d\phi^{2} \right)\, , \label{gansatz} \\
f_{\mu \nu} dx^{\mu} dx^{\nu}
=& e^{2\rho_{2}(t,x)} \left( -dt^{2} + dx^{2} \right) \nonumber \\
&+ e^{-2\varphi_{2}(t,x)}\left( d\theta^{2} + \sin^{2} \theta d\phi^{2} \right)\, . \label{fansatz}
\end{align}
Here, two-dimensional metric, corresponding to $t$ and $x$ components, is written in the conformal 
gauge and $x$ is the coordinate on the one-sphere and has the period of $2\pi$.
We should also note that the black hole and cosmological horizons are located at 
same place \cite{Deffayet:2011rh}, respectively.

Under the above ansatz, we can calculate each components of the Einstein tensor 
$G_{\mu \nu} \equiv R_{\mu \nu} - \frac{1}{2}Rg_{\mu \nu}$,
\begin{align}
&G_{tt} =  \dot{\varphi}^{2} - 2 \dot{\rho}\dot{\varphi} - 2\rho^{\prime}\varphi^{\prime} 
+ 2\varphi^{\prime \prime}  - 3{\varphi^{\prime}}^{2} + e^{2(\varphi + \rho)} \, ,\nonumber \\
&G_{tx} = 2\dot{\varphi}^{\prime} - 2\varphi^{\prime}\dot{\varphi} - 2\rho^{\prime}\dot{\varphi} 
 - 2\dot{\rho}\varphi^{\prime} \, , \nonumber \\
&G_{xx} = {\varphi^{\prime}}^{2} - 2 \dot{\rho}\dot{\varphi} - 2\rho^{\prime}\varphi^{\prime} 
+ 2\ddot{\varphi} - 3\dot{\varphi}^{2}  - e^{2(\varphi + \rho)} \, ,\nonumber \\
&G_{\theta \theta} = e^{-2(\rho + \varphi)} \left(  -\ddot{\rho} + \rho^{\prime \prime} 
+ \ddot{\varphi} - \varphi^{\prime \prime} - \dot{\varphi}^{2} +{\varphi^{\prime}}^{2} \right) \, , 
\nonumber \\
&G_{\phi \phi} = e^{-2(\rho + \varphi)} \left(  -\ddot{\rho} + \rho^{\prime \prime} + \ddot{\varphi} 
 - \varphi^{\prime \prime} - \dot{\varphi}^{2} +{\varphi^{\prime}}^{2} \right) \sin^{2} \theta \, .
\label{Einstein tensors}
\end{align}
Here, $\dot{} \equiv \partial /\partial t$ and ${}^{\prime} \equiv \partial / \partial x$,
and sub indices are omitted for simplicity.

Furthermore, we need to calculate the interaction terms in Eqs.~(\ref{geq}) and (\ref{feq}). 
The interaction terms in each equations of motion are written in terms of $\sqrt{g^{-1}f}$ and 
$\sqrt{f^{-1}g}$.
Defining $\mathbf{A}=\sqrt{g^{-1}f}$ and $\mathbf{B}=\sqrt{f^{-1}g}$ for convention, 
these two matrices are expressed as follows:
\begin{align}
&\mathbf{A} = \mathrm{diag} \left( e^{-\zeta}, e^{-\zeta}, e^{\xi}, e^{\xi} \right)\, , \\
&\mathbf{B} = \mathrm{diag} \left( e^{\zeta}, e^{\zeta}, e^{-\xi}, e^{-\xi} \right) \,,
\end{align}
where we define $ \zeta \equiv \rho_{1}-\rho_{2}, \xi \equiv \varphi_{1} - \varphi_{2}$.

After short calculation, we obtain $Y_{n}$s which take the following forms:
\begin{align}
Y_{0}(\mathbf{A}) 
=& \mathbf{1}\, , \nonumber \\
Y_{1}(\mathbf{A})
=& \mathrm{diag} \left( -e^{-\zeta} - 2e^{\xi}, -e^{-\zeta} - 2e^{\xi}\, , \right. \nonumber \\
& \hphantom{diag ( -e^{-\zeta}  } 
\left. - 2e^{-\zeta} - e^{\xi}, - 2e^{-\zeta} - e^{\xi} \right)\, , \nonumber \\
Y_{2}(\mathbf{A}) 
=& \mathrm{diag} \left(  2e^{-\zeta + \xi} + e^{2\xi}, 2e^{-\zeta + \xi} + e^{2\xi}\, , \right. 
\nonumber \\
& \hphantom{diag (  2e }
\left. e^{-2\zeta} + 2e^{-\zeta + \xi}, e^{-2\zeta} + 2e^{-\zeta + \xi} \right)\, , \nonumber \\
Y_{3}(\mathbf{A})
=& \mathrm{diag} \left( -e^{-\zeta + 2 \xi}, -e^{-\zeta + 2\xi}\, , \right. \nonumber \\
& \hphantom{\mathrm{diag} ( -e^{-\zeta + 2 \xi}, }
\left. - e^{-2\zeta + \xi}, -e^{-2\zeta + \xi} \right) 
\label{interaction gf}
\end{align}
\begin{align}
Y_{0}(\mathbf{B}) =& \mathbf{1}\, , \nonumber \\
Y_{1}(\mathbf{B})
=& \mathrm{diag} \left( -e^{\zeta} - 2e^{-\xi}, -e^{\zeta} - 2e^{-\xi}\, , \right. \nonumber \\
&  \hphantom{diag ( -e^{\zeta} }
\left. - 2e^{\zeta} - e^{-\xi}, - 2e^{\zeta} - e^{-\xi} \right)\, , \nonumber \\
Y_{2}(\mathbf{B}) 
=& \mathrm{diag} \left( 2e^{\zeta - \xi} + e^{ -2\xi}, 2e^{\zeta - \xi} + e^{ -2\xi}\, , 
\right. \nonumber \\
& \hphantom{diag ( 2e }
\left. e^{2\zeta} + 2e^{\zeta - \xi}, e^{2\zeta} + 2e^{\zeta - \xi} \right) \, , \nonumber \\
Y_{3}(\mathbf{B})
=& \mathrm{diag} \left( -e^{\zeta- 2 \xi}, -e^{\zeta - 2\xi}\, , \right. \nonumber \\
& \hphantom{\mathrm{diag} ( -e^{\zeta- 2 \xi}, }
\left. - e^{2\zeta - \xi}, -e^{2\zeta - \xi} \right)\, .
\label{interaction fg}
\end{align}
Next, we consider the equations of motion.
When we choose $M_{g}=M_{f}$, the effective Planck mass scale is given by
\begin{align}
M^{2}_\mathrm{eff} = \frac{1}{2}M^{2}_{g} = \frac{1}{2}M^{2}_{f}\, .
\end{align}
Now we do not restrict the combinations of parameters in the interaction terms
but consider general case, that is, the case that there are independent five parameters 
$\beta_{n}$s.
Two equations of motion for $g_{\mu \nu}$ and $f_{\mu \nu}$ take the following form:
\begin{align}
0 =& R_{\mu \nu}(g) - \frac{1}{2}R(g)g_{\mu \nu} \nonumber \\
& + \frac{1}{2} m_{0}^{2} 
\left[ \beta_{0} Y^{\lambda}_{(0) \nu}(\mathbf{A}) - \beta_{1} Y^{\lambda}_{(1) \nu}(\mathbf{A}) 
\right. \nonumber \\
& \hphantom{+ \frac{1}{2} m_{0}^{2} [ \beta_{0} } \left.
+ \beta_{2} Y^{\lambda}_{(2) \nu}(\mathbf{A}) - \beta_{3} Y^{\lambda}_{(3) \nu}(\mathbf{A}) \right] 
g_{\mu \lambda}\, , \\
0 =& R_{\mu \nu}(f) - \frac{1}{2}R(f)f_{\mu \nu} \nonumber \\
& + \frac{1}{2} m_{0}^{2} 
\left[ \beta_{4} Y^{\lambda}_{(0) \nu}(\mathbf{B}) - \beta_{3} Y^{\lambda}_{(1) \nu}(\mathbf{B}) 
\right. \nonumber \\
& \hphantom{+ \frac{1}{2} m_{0}^{2} [ \beta_{4}  } \left.
+ \beta_{2} Y^{\lambda}_{(2) \nu}(\mathbf{B}) - \beta_{1} Y^{\lambda}_{(3) \nu}(\mathbf{B}) \right] 
f_{\mu \lambda}\, .
\end{align}
As we have discussed, one can show 
that we obtain asymptotically de Sitter solution for the specific combinations of parameters
under the ansatz (\ref{proportional relation}).
Furthermore, when we impose the spherical symmetry on the solutions,
the Schwarzschild-de Sitter space-time can be solutions in our setting.
So, we will consider the condition to obtain the Nariai space-time as a background solution.

In the coordinate system of Eqs~.(\ref{gansatz}) and (\ref{fansatz}), 
the Nariai solutions are expressed as follows:
\begin{align}
&g_{\mu \nu}dx^{\mu} dx^{\nu} = 
\frac{1}{\Lambda \cos^{2}t}\left( -dt^{2} + dx^{2} \right)
+ \frac{1}{\Lambda}d\Omega^{2}\, , \\
&f_{\mu \nu} dx^{\mu} dx^{\nu} = 
\frac{C^{2}}{\Lambda \cos^{2}t} \left( -dt^{2} + dx^{2} \right) 
+ \frac{C^{2}}{\Lambda} d\Omega^{2}\, .
\end{align}
Therefore, the corresponding $\rho(t,x)$s and $\varphi(t,x)$s in the Nariai solutions 
are given by
\begin{align}
&e^{2\rho_{1}(t,x)} = \frac{1}{\Lambda \cos^{2}t} , \quad  e^{-2\varphi_{1}(t,x)} 
= \frac{1}{\Lambda}\, , \nonumber \\
&e^{2\rho_{2}(t,x)} = \frac{C^{2}}{\Lambda \cos^{2}t}\, , 
\quad  e^{-2\varphi_{2}(t,x)} = \frac{C^{2}}{\Lambda}\, ,
\end{align}
that is,
\begin{align}
&\rho_{1} = - \frac{1}{2} \mathrm{log} \, \Lambda-\mathrm{log} (\cos t) \, ,\nonumber \\
&\varphi_{1} = \frac{1}{2} \mathrm{log} \, \Lambda\, , \nonumber \\
&\rho_{2}= \mathrm{log} \,C - \frac{1}{2} \mathrm{log} \, \Lambda -\mathrm{log} (\cos t) \, ,
\nonumber \\
&\varphi_{2} = -\mathrm{log} \, C + \frac{1}{2}\mathrm{log} \, \Lambda\, , \nonumber \\
&\zeta = - \mathrm{log} \, C , \quad \xi = \mathrm{log} \, C \, .\label{Nariai components}
\end{align}
Now, we substitute these solutions into the equations of motion.
Substituting (\ref{Nariai components}) into (\ref{Einstein tensors}), (\ref{interaction gf}), and 
(\ref{interaction fg}),
the Einstein tensors and the interaction terms take the following forms:
\begin{align}
&G_{tt}(g) = G_{tt}(f) = \frac{1}{\cos^{2}t} \, , \nonumber \\
&G_{tx}(g) = G_{tx}(f) = 0 \, , \nonumber \\
&G_{xx}(g) = G_{xx}(f) = -\frac{1}{\cos^{2}t} \, , \nonumber \\
&G_{\theta \theta}(g) = G_{\theta \theta}(f) = -1 , \nonumber \\
&G_{\phi \phi}(g) = G_{\phi \phi}(f) = -\sin^{2} \theta \, , \label{Nariai Einstein tensors}
\end{align}
\begin{align}
&Y_{0}(\mathbf{A}) = \mathbf{1}\, , \quad Y_{1}(\mathbf{A}) = -3C \mathbf{1}\, , \nonumber \\
&Y_{2}(\mathbf{A}) = 3C^{2} \mathbf{1}\, , \quad Y_{3}(\mathbf{A}) = -C^{3} \mathbf{1} \, ,
\end{align}
\begin{align}
&Y_{0}(\mathbf{B}) = \mathbf{1}\, , \quad Y_{1}(\mathbf{B}) =  -3C^{-1} \mathbf{1}\, , \nonumber \\
&Y_{2}(\mathbf{B}) = 3C^{-2} \mathbf{1}\, , \quad Y_{3}(\mathbf{B}) = -C^{-3} \mathbf{1}\, .
\end{align}
And, we find that two equations of motion are given by
\begin{align}
0 =& 1 - \frac{1}{2\Lambda} m_{0}^{2} \left[ \beta_{0} +3\beta_{1}C + 3\beta_{2}C^{2}  
+ \beta_{3}C^{3} \right] \, , \\
0 =& 1 - \frac{C^{2}}{2\Lambda} m_{0}^{2} \left[ \beta_{4}  + 3\beta_{3}C^{-1} + 3\beta_{2} C^{-2} 
+ \beta_{1} C^{-3} \right]\, . 
\end{align}
Here, one can identify the two cosmological constants as follows:
\begin{align}
&\Lambda_{g}(C) = \Lambda 
= \frac{1}{2} m_{0}^{2} \left[ \beta_{0} +3\beta_{1}C + 3\beta_{2}C^{2}  
+ \beta_{3}C^{3} \right]\, , \label{lambdag}\\
&\Lambda_{f}(C) = \frac{1}{2} m_{0}^{2} \left[ \beta_{4}  + 3\beta_{3}C^{-1} + 3\beta_{2} C^{-2} 
+ \beta_{1} C^{-3} \right] \label{lambdaf}\, .
\end{align}
Then we obtain the quartic equation of $C$, 
\begin{align}
\Lambda_{g}(C)=C^{2}\Lambda_{f}(C) \, , \label{Ceq}
\end{align}
by using the two equations of motion.

When one choose the interacting parameters $\beta_{n}$s, 
Eq.~(\ref{Ceq}) are determined and the consistent $C$ can be given as a solution
which reproduces the positive cosmological constants in Eqs.~(\ref{lambdag}) and (\ref{lambdaf}).
Therefore, in order to obtain the Nariai solutions, 
all we have to do is to find the suitable interacting parameters.
In the following discussion, we assume that the $\beta_{n}$s are chosen to realize 
the asymptotically de Sitter space-time.

\subsection{Perturbations}

Next, we define the perturbations as follows:
\begin{align}
&\rho_{1} \equiv \bar{\rho}_{1} + \delta \rho_{1}(t,x)\, , \quad 
\varphi_{1} \equiv \bar{\varphi}_{1} + \delta \varphi_{1}(t,x)\, , \nonumber \\
&\rho_{2} \equiv \bar{\rho}_{2} + \delta \rho_{2}(t,x)\, , \quad 
\varphi_{2} \equiv \bar{\varphi}_{2} + \delta  \varphi_{2}(t,x)\, .
\end{align}
Here, $\bar{\rho}$s and $\bar{\varphi}$s correspond to the unperturbed Nariai space-time and
$\delta \rho$s and $\delta \varphi$s are the perturbations.
By substituting the above expressions into (\ref{gansatz}) and (\ref{fansatz}),
we find the metric perturbations of $g_{\mu \nu}$ and $f_{\mu \nu}$ in the first order,
\begin{align}
g_{\mu \nu} =& 
\mathrm{diag} \left( -e^{2\bar{\rho}_{1}}, e^{2\bar{\rho}_{1}}\, , 
e^{-2\bar{\varphi}_{1}}, e^{-2\bar{\varphi}_{1}} \sin^{2}\theta  \right) \nonumber \\
&+ \mathrm{diag} \left( -2e^{2\bar{\rho}_{1}} \delta\rho_{1} , 
2e^{2\bar{\rho}_{1}} \delta\rho_{1} , 
\right. \nonumber \\
& \hphantom{\mathrm{diag} ( -e^{2\bar{\rho}_{1}}} \left.
-2e^{-2\bar{\varphi}_{1}} \delta\varphi_{1}, 
-2e^{-2\bar{\varphi}_{1}} \delta\varphi_{1} \sin^{2}\theta  \right) \nonumber \\
=& \bar{g}_{\mu \nu} + \delta g_{\mu \nu} \, ,
\end{align}
\begin{align}
f_{\mu \nu} =& 
\mathrm{diag} \left( -e^{2\bar{\rho}_{2}}, e^{2\bar{\rho}_{2}}, 
e^{-2\bar{\varphi}_{2}}, e^{-2\bar{\varphi}_{2}} \sin^{2}\theta  \right) \nonumber \\
&+ \mathrm{diag} \left( -2e^{2\rho_{2}} \delta\rho_{2} , 2e^{2\rho_{2}} \delta\rho_{2}\, , 
\right. \nonumber \\
& \hphantom{\mathrm{diag} ( -e^{2\bar{\rho}_{2}},} \left.
 -2e^{-2\varphi_{2}} \delta\varphi_{2}, 
 -2e^{-2\varphi_{2}} \delta\varphi_{2} \sin^{2}\theta  \right) \nonumber \\
=& \bar{f}_{\mu \nu} + \delta f_{\mu \nu}\, , 
\end{align}
where, we define 
\begin{align}
\delta g_{\mu \nu}
=& \mathrm{diag} \left( -\frac{2}{\Lambda \cos^{2}t}\delta\rho_{1}\, , 
\frac{2}{\Lambda \cos^{2}t}\delta\rho_{1}\, , \right. \nonumber \\
&\hphantom{\mathrm{diag} ( -\frac{2}{\Lambda \cos^{2}t}\delta\rho_{1}} \left.
-\frac{2}{\Lambda}\delta\varphi_{1}, -\frac{2}{\Lambda} \delta\varphi_{1} \sin^{2}\theta \right)\, , 
\end{align}
\begin{align}
\delta f_{\mu \nu}
=& \mathrm{diag} \left( -\frac{2C^{2}}{\Lambda \cos^{2}t}\delta\rho_{1}\, , 
\frac{2C^{2}}{\Lambda \cos^{2}t}\delta\rho_{1}\, , \right. \nonumber \\
& \hphantom{\mathrm{diag} ( -\frac{2C^{2}}{\Lambda \cos^{2}t}} \left.
-\frac{2C^{2}}{\Lambda}\delta\varphi_{1}, -\frac{2C^{2}}{\Lambda} \delta\varphi_{1} 
\sin^{2}\theta  \right) \, .
\end{align}
We now evaluate the equations of motion for the perturbation.
At first, we calculate the perturbation of the Einstein tensor in the first order.
When we substitute the metric perturbations into (\ref{Einstein tensors}),
we obtain the deviations of $G_{\mu \nu}(g)$, 
\begin{align}
G_{tt}(g) 
=& \frac{1}{\cos^{2}t}+ 2\delta\varphi_{1}^{\prime \prime} \nonumber \\
& -2\tan t \delta \dot{\varphi_{1}} + \frac{2}{\cos^{2}t}(\delta \varphi_{1} + \delta \rho_{1}) \, ,\\
G_{tx}(g) 
=& 2\delta \dot{\varphi}^{\prime}_{1} - 2\tan t \delta \varphi_{1}^{\prime} \, ,\\
G_{xx}(g) 
=& - \frac{1}{\cos^{2}t} + 2\delta \ddot{\varphi_{1}} \nonumber \\
&-2\tan t \delta \dot{\varphi_{1}} - \frac{2}{\cos^{2}t} (\delta \varphi_{1} + \delta \rho_{1}) \, ,\\
G_{\theta \theta}(g) 
=& -1 +2(\delta \rho_{1} + \delta \varphi_{1}) \nonumber \\
& +\cos^{2}t \left(- \delta \ddot{\rho_{1}} + \delta \rho_{1}^{\prime \prime} 
+ \delta \ddot{\varphi_{1}} - \delta \varphi_{1}^{\prime \prime}\right) \, ,\\
G_{\phi \phi}(g) 
=& -\sin^{2} \theta + \sin^{2} \theta 
\left \{ 2(\delta \rho_{1} + \delta \varphi_{1}) \right. \nonumber \\
& \left. +\cos^{2}t \left(- \delta \ddot{\rho_{1}} + \delta \rho_{1}^{\prime \prime} 
+ \delta \ddot{\varphi_{1}} - \delta \varphi_{1}^{\prime \prime}\right) \right \} \, .
\end{align}
Then, we define the deviations of the Einstein tensor from the Nariai space-time.
Note that the deviations of the Einstein tensor for $f_{\mu \nu}$ are obtained by changing 
$\rho_{1} \rightarrow \rho_{2}$ and
$\varphi_{1} \rightarrow \varphi_{2}$, 
because $\rho_{1}+\varphi_{1} = \rho_{2} + \varphi_{2}$ 
and $\mathrm{log} \, C$ is constant.
Compared with Eq.~(\ref{Nariai Einstein tensors}), 
we find the deviations of the Einstein tensor are given by
\begin{align}
\delta G_{tt}
=& 
2\delta\varphi^{\prime \prime} -2\tan t \delta \dot{\varphi} 
+ \frac{2}{\cos^{2}t}(\delta \varphi + \delta \rho) \, , \\
\delta G_{tx}
=&
2\delta \dot{\varphi}^{\prime} - 2\tan t \delta \varphi^{\prime} \, ,\\
\delta G_{xx}
=&
2\delta \ddot{\varphi} -2\tan t \delta \dot{\varphi} 
- \frac{2}{\cos^{2}t} (\delta \varphi + \delta \rho) \, ,\\
\delta G_{\theta \theta}
=&
2(\delta \rho + \delta \varphi)  \nonumber \\
&+\cos^{2}t \left(- \delta \ddot{\rho} + \delta \rho^{\prime \prime} 
+ \delta \ddot{\varphi} - \delta \varphi^{\prime \prime}\right) \, ,\\
\delta G_{\phi \phi}
=&
\sin^{2} \theta  \left \{ 2(\delta \rho + \delta \varphi) \right. \nonumber \\
& \left. 
+\cos^{2}t \left(- \delta \ddot{\rho} + \delta \rho^{\prime \prime} 
+ \delta \ddot{\varphi} - \delta \varphi^{\prime \prime}\right) \right \} \, .
\end{align}
Next, we evaluate the interaction terms.
We define the deviation of $\zeta$ and $\xi$ as follows:
\begin{align}
&\zeta = - \mathrm{log} \, C + \delta \rho_{1} - \delta \rho_{2} 
\equiv  \bar{\zeta} + \delta \zeta\, , \\
&\xi = \mathrm{log} \, C+ \delta \varphi_{1} - \delta \varphi_{2} 
\equiv \bar{\xi} + \delta \xi\, .
\end{align}
Then, we can calculate the deviations of the interaction terms $Y_{n}$s 
from the Nariai space-time, and they are given by
\begin{align}
&\delta Y_{0}(\mathbf{A}) = \mathbf{0}\, , \quad \delta Y_{1}(\mathbf{A}) = - C^{-1} \mathbf{Z} 
\, ,\nonumber \\
&\delta Y_{2}(\mathbf{A}) =2C^{-2} \mathbf{Z}\, , \quad \delta Y_{3}(\mathbf{A}) = - C^{-3} 
\mathbf{Z}\, , 
\end{align}
\begin{align}
&\delta Y_{0}(\mathbf{B}) = \mathbf{0}\, , \quad \delta Y_{1}(\mathbf{B}) = C \mathbf{Z} \, ,
\nonumber \\
&\delta Y_{2}(\mathbf{B}) = -2 C^{2} \mathbf{Z}\, , 
\quad \delta Y_{3}(\mathbf{B}) = C^{3} \mathbf{Z}\, , 
\end{align}
where we define the tensor $\mathbf{Z}$ as follows,
\begin{align}
\mathbf{Z} = \mathrm{diag} \left( \delta \zeta - 2\delta \xi, \delta \zeta - 2\delta \xi, 
2\delta \zeta - \delta \xi, 2\delta \zeta - \delta \xi \right)\, .
\end{align}
Finally, we consider the equations for the perturbations.
For the convention, we express the equations of motion as follows:
\begin{align}
G_{\mu \nu}(g) + I^{\lambda}_{\ \nu}(\mathbf{A})g_{\mu \lambda}=0\, , \\
G_{\mu \nu}(f) + I^{\lambda}_{\ \nu}(\mathbf{B})f_{\mu \lambda}=0\, , 
\end{align}
where $I^{\lambda}_{\ \nu}$s are the sum of $Y_{n}$s.
When we consider the perturbation up to first order, above equations are divided by background 
part and deviation part, and the equations for the deviation take the following forms:
\begin{align}
\delta G_{\mu \nu}(g) + \delta I^{\lambda}_{\ \nu}(\mathbf{A})g_{\mu \lambda} 
+ I^{\lambda}_{\ \nu}(\mathbf{B}) \delta g_{\mu \lambda}=0\, , 
\label{gperturb}\\
\delta G_{\mu \nu}(f) + \delta I^{\lambda}_{\ \nu}(\mathbf{B})f_{\mu \lambda} 
+ I^{\lambda}_{\ \nu}(\mathbf{A})\delta f_{\mu \lambda} =0\, .
\label{fperturb}
\end{align}
Here, we define
\begin{align}
I(\mathbf{A}) 
=& \frac{1}{2} m_{0}^{2} \left[ \beta_{0} +3\beta_{1}C + 3\beta_{2}C^{2}  
+ \beta_{3}C^{3} \right] \mathbf{1} \nonumber \\
=& \Lambda \mathbf{1}\, , \\ 
I(\mathbf{B}) 
=& \frac{1}{2} m_{0}^{2} \left[ \beta_{4} +3\beta_{3}C^{-1} 
+3\beta_{2} C^{-2} + \beta_{1} C^{-3} \right] \mathbf{1} \nonumber \\
=& \frac{\Lambda}{C^{2}} \mathbf{1} \, , \\
\delta I(\mathbf{A}) 
=& - \frac{1}{2} m_{0}^{2} \left[ \beta_{1}C +2\beta_{2}C^{2} 
+\beta_{3}C^{3} \right] \mathbf{Z} \nonumber \\
=& - C_{1} \mathbf{Z} \, , \label{nariai int a} \\
\delta I(\mathbf{B}) 
=& \frac{1}{2} m_{0}^{2} \left[ \beta_{3}C^{-1} +2\beta_{2} C^{-2} 
+ \beta_{1} C^{-3} \right] \mathbf{Z} \nonumber \\
=& C^{-4}C_{1}\mathbf{Z}\, , \\
C_{1} \equiv& \frac{1}{2} m_{0}^{2} \left[ \beta_{4} +3\beta_{3}C^{-1} 
+3\beta_{2} C^{-2} + \beta_{1} C^{-3} \right] \, . \label{nariai int b}
\end{align}

\subsection{Evolution of black hole horizon}

In order to describe the evolution of black holes due to the perturbations,
we need to know where the horizons are located for $g_{\mu \nu}$ and $f_{\mu \nu}$.
In the following, we consider the black hole horizon for $g_{\mu \nu}$ at first.
Let us specify the form of perturbations 
so that the two-sphere radius $e^{-\varphi_{1}}$ varies along the one-sphere coordinate $x$:
\begin{align}
e^{2\varphi_{1}} = \Lambda \left \{1 + 2\epsilon \sigma_{1}(t) \cos x \right \}\, , \quad
|\epsilon| \ll 1\, , 
\end{align}
that is,
\begin{align}
\delta \varphi_{1} \equiv \epsilon \sigma_{1}(t) \cos x \, . 
\label{gperturb2}
\end{align}
Note that above form of perturbation is consistent with Eq.~(\ref{gperturb}).
Substituting the above form of perturbation into the $(t,x)$ component of (\ref{gperturb}),
we obtain
\begin{align}
\dot{\sigma_{1}} = \sigma_{1}\tan t\, .
\end{align}
With the boundary condition, $\dot{\sigma_{1}}=0$ at $t=0$, the solution is
\begin{align}
\sigma_{1}(t) = \frac{\sigma_{g}}{\cos t}\, . \label{classical sigma}
\end{align} 
For this solution, we consider the time evolution of the black hole horizon.

The condition for a horizon is $ (\nabla \delta \varphi_{1})^{2} = 0 $, 
which is required that the gradient of the two-sphere size is null.
Here, Eq.~(\ref{gperturb2}) yields
\begin{align}
\delta \dot{\varphi_{1}} = \epsilon \dot{\sigma_{1}} \cos x\, , \quad
\delta \varphi^{\prime}_{1} = - \epsilon \sigma_{1} \sin x\, .
\end{align}
From above conditions, 
locations of the black hole horizon $x_{b}$ and cosmological horizon $x_{c}$ are defined as follows:
\begin{align}
x_{b} = \arctan \left| \frac{\dot{\sigma}}{\sigma} \right|\, , \quad x_{c}=\pi - x_{b}\, .
\end{align}
Therefore, the size of the black hole horizon, $r_{b}$, is given by
\begin{align}
r_{b}^{-2}(t) = e^{2\varphi(t,x_{b})} = \Lambda \left\{ 1 + 2 \epsilon \delta(t) \right\}\, ,
\end{align}
where the we define the perturbation for the horizon $\delta(t)$,
\begin{align}
\delta(t) \equiv \sigma_{1}(t) \cos x_{b} = \sigma_{1} \left \{ 1 
+ \left( \frac{\dot{\sigma_{1}}}{\sigma_{1}} \right)^{2} \right \}^{-1/2} \, .\label{horizon}
\end{align}
Then, substituting Eq.~(\ref{classical sigma}) into Eq.~(\ref{horizon}), we obtain
\begin{align}
\delta(t) = \sigma_{g} = \mbox{const}\, .
\end{align}
This means that no anti-evaporation takes place 
and horizon size remains that of the initial perturbation.
This is just a static Schwarzschild-de Sitter black holes of nearly maximal mass. 

Note that, if we define the same form of perturbation for $\delta \varphi_{2}$ as that for 
$\delta \varphi_{1}$,
we obtain same results because the equations have same form as that of $\varphi_{1}$.
Then, one can find that anti-evaporation does not occur for two metrics $g_{\mu \nu}$ 
and $f_{\mu \nu}$ on classical level.

\section{Difference from GR}

In the previous section, we found that the anti-evaporation is not realized in bigravity on classical 
level, which is not changed from the result in the general relativity.
In this section, we focus on the problem how we can identify the difference between the case in 
general relativity and in bigravity.

When we substitute the perturbations into the $(t,t)$ and $(x,x)$ components of (\ref{gperturb}), we 
obtain
\begin{align}
\delta \zeta - 2\delta \xi = 0\, .
\end{align}
Thus, deviations of the interaction terms (\ref{nariai int a}), (\ref{nariai int b}) vanish 
if $\delta \zeta =0$ or $\delta \xi = 0$.
When we define the perturbation for $f_{\mu \nu}$ as 
\begin{align}
e^{2\varphi_{2}} = \frac{\Lambda}{C^{2}} \left \{1 + 2\epsilon \sigma_{2}(t) \cos x \right \}\, , 
\quad \sigma_{2}(t) = \frac{\sigma_{f}}{\cos t}\, ,
\end{align}
$\delta \xi$ vanishes in the case where the amplitude of the perturbations are identical, 
$\sigma_{g}=\sigma_{f}$.
This means that the two sets of metric perturbations are proportional to each other and 
the relation between the perturbations is not changed from the background,
$\delta f_{\mu \nu} = C^{2} \delta g_{\mu \nu}$.
In this case, whole metrics including the perturbations are proportional and it does not lead to 
difference from general relativity.
Therefore, we cannot distinguish bigravity theory from general relativity.

Note that, regarding the perturbation, one can introduce the different forms between 
$\delta \varphi_{1}$ and $\delta \varphi_{2}$.
For instance, we may assume the following form,
\begin{align}
& \delta \varphi_{1} = \epsilon \frac{\sigma_{g}}{\cos t} \cos (x + \alpha) \, ,\\
& \delta \varphi_{2} = \epsilon \frac{\sigma_{f}}{\cos t} \cos (x + \beta) \, ,
\end{align}
which is consistent with the equations Eqs.~(\ref{gperturb}) and (\ref{fperturb}).
In this case, the amplitude and phase can take independent values
and these are difference from the general relativity.

\section{Summary and Discussion}

We have studied the possibility of the anti-evaporation on classical level in the bigravity.
For the assumption $f_{\mu \nu} = C^{2} g_{\mu \nu}$ 
and particular parameters $\beta_{n}$s 
and the Planck mass scales $M_{g} = M_{f}$, 
we obtained the asymptotically de Sitter space-time.
When we considered the perturbations around the Nariai space-time,
the size of black hole horizon does not increase.
And we have found that the anti-evaporation does not take place on the classical level
although the equations of motion are different from general relativity.

%%%%%%%%%%
When we assume the perturbation (\ref{gperturb2}),
Eq.~(\ref{classical sigma}) is derived from the $(t,x)$ component of the Eq.~(\ref{gperturb}).
However, the non-diagonal components of the Eq.~(\ref{gperturb}) take the forms identical with 
that in general relativity
because the interaction terms do not modify the non-diagonal components.
Therefore, the result that the size of black hole horizon does not increase 
is not changed from that in the general relativity.
In the $F(R)$ gravity, however, the anti-evaporation can occur 
because the equations of motion are modified in different way.
%%%%%%%%%%%%%%

In order to realize the anti-evaporation, 
we need to take the quantum corrections into account 
%in order to realize the anti-evaporation
in a similar way to general relativity.
The effective action due to the quantum correction of matter fields 
is given in the same manner in general relativity.
Interesting problem is probably to study if we need to introduce the quantum corrections
for only one metric or both metrics.
For instance, when one regards $g_{\mu \nu}$ as the metric which describes our world 
and finds that the anti-evaporation occurs with introducing the quantum corrections
only to $f_{\mu \nu}$ sector,
the black holes grow although our world is exactly classical.

There is another way to realize the anti-evaporation by modification to $F(R)$ bigravity theory
\cite{Nojiri:2012re,Nojiri:2012zu,Kluson:2013yaa,Bamba:2013hza}.
This theory modify the kinetic terms of bigravity, from the Ricci scalar to the function of it. 
In $F(R)$ bigravity, we find similar problem to introducing the quantum corrections.
That is, we need to study if the modification is required for only one metric or both metrics.

\section*{Acknowledgements}

The authors are deeply indebted to Sergei.~D.~Odintsov for constructive advices and discussion in 
the early stage of this research.
T.K is partially supported by the Nagoya University Program for Leading Graduate Schools funded 
by the Ministry of Education of the Japanese Government under the program number N01.
This work is also supported by the JSPS Grant-in-Aid for Scientific 
Research (S) \# 22224003 and (C) \# 23540296 (S.N.). 

\appendix

\section{Geometrical quantities}

The connections and curvature in the conformal gauge are given as follows:
\begin{align}
&\Gamma^{t}_{\ tt} = \Gamma^{t}_{\ xx} = \dot{\rho}, \quad \Gamma^{t}_{\ tx} = \rho^{\prime}\, , 
\nonumber \\
&\Gamma^{t}_{\ \theta \theta} = - \dot{\varphi} e^{-2(\rho + \varphi)}\, , \quad
\Gamma^{t}_{\ \phi \phi} = - \dot{\varphi} e^{-2(\rho + \varphi)} \sin^{2} \theta\, ,  \nonumber \\
&\Gamma^{x}_{\ tx} = \dot{\rho} , \quad \Gamma^{x}_{\ xx} = \Gamma^{x}_{\ tt}  = \rho^{\prime}\, , 
\nonumber \\
&\Gamma^{x}_{\ \theta \theta} = \varphi^{\prime} e^{-2(\rho + \varphi)}\, , \quad 
\Gamma^{x}_{\ \phi \phi} = \varphi^{\prime} e^{-2(\rho + \varphi)} \sin^{2} \theta, \nonumber \\
&\Gamma^{\theta}_{\ t \theta} = - \dot{\varphi}\, , \quad 
\Gamma^{\theta}_{\ x \theta} = - \varphi^{\prime}\, , \quad
\Gamma^{\theta}_{\ \phi \phi} = - \sin \theta \cos \theta \, , \nonumber \\
&\Gamma^{\phi}_{\ t \phi} = - \dot{\varphi}, \quad \Gamma^{\phi}_{\ x \phi} = - \varphi^{\prime}\, , 
\quad
\Gamma^{\phi}_{\ \theta \phi} = \cot \theta\, , \nonumber 
\end{align}
\begin{align}
&R_{tt} = - \ddot{\rho} + 2 \ddot{\varphi} 
+ \rho^{\prime \prime} - 2\dot{\varphi}^{2} - 2 \dot{\rho}\dot{\varphi} 
 - 2\rho^{\prime}\varphi^{\prime}\, , \nonumber \\
&R_{xx} = \ddot{\rho} + 2\varphi^{\prime \prime} - \rho^{\prime \prime} - 2{\varphi^{\prime}}^{2} 
 - 2 \dot{\rho}\dot{\varphi} - 2\rho^{\prime}\varphi^{\prime}\, , \nonumber \\
&R_{tx} = 2\dot{\varphi}^{\prime} - 2\varphi^{\prime}\dot{\varphi} - 2\rho^{\prime}\dot{\varphi} 
 - 2\dot{\rho}\varphi^{\prime}\, , \nonumber \\
&R_{\theta \theta} = 1 + e^{-2(\rho + \varphi)} \left( - \ddot{\varphi} + \varphi^{\prime \prime} 
+ 2\dot{\varphi}^{2} - 2{\varphi^{\prime}}^{2} \right) \, , \nonumber \\
&R_{\phi \phi} = \left \{ 1 + e^{-2(\rho + \varphi)} \left( - \ddot{\varphi} + \varphi^{\prime \prime} 
+ 2\dot{\varphi}^{2} - 2{\varphi^{\prime}}^{2} \right) \right \} \sin^{2} \theta\, , 
\nonumber 
\end{align}
\begin{align}
&R = \left( 2\ddot{\rho} - 2\rho^{\prime \prime} - 4\ddot{\varphi} + 4\varphi^{\prime \prime} 
+ 6\dot{\varphi}^{2} - 6{\varphi^{\prime}}^{2} \right)e^{-2\rho} + 2e^{2\varphi}\, .
\nonumber
\end{align}
Here, $\dot{} \equiv \partial /\partial t$ and ${}^{\prime} \equiv \partial / \partial x$.

\section{Perturbations}

The equations for the perturbations are given by as follows:
\begin{itemize}
\item $(t,t)$ component of (\ref{gperturb}) 
\begin{align}
0=&\delta\varphi_{1}^{\prime \prime} - \tan t \delta \dot{\varphi_{1}} 
+ \frac{1}{\cos^{2}t} \delta \varphi_{1}  \nonumber \\
&+ \frac{C_{1}}{2\Lambda \cos^{2} t}  (\delta \zeta - 2\delta \xi) \, . \label{gtt}
\end{align}
\item $(t,x)$ component of (\ref{gperturb}) 
\begin{align}
0=\delta \dot{\varphi}^{\prime}_{1} - \tan t \delta \varphi_{1}^{\prime} \, .\label{gtx}
\end{align}
\item $(x,x)$ component of (\ref{gperturb}) 
\begin{align}
0=&\delta \ddot{\varphi_{1}} -\tan t \delta \dot{\varphi_{1}} 
- \frac{1}{\cos^{2}t} \delta \varphi_{1} \nonumber \\
& - \frac{C_{1}}{2\Lambda \cos^{2}t}  (\delta \zeta - 2\delta \xi) \, .
\label{gxx}
\end{align}
\item $(\theta, \theta)$, $(\phi, \phi)$ component of (\ref{gperturb}) 
\begin{align}
0=&2\delta \rho_{1} 
+\cos^{2}t \left(- \delta \ddot{\rho_{1}} + \delta \rho_{1}^{\prime \prime} 
+ \delta \ddot{\varphi_{1}} 
- \delta \varphi_{1}^{\prime \prime}\right) \nonumber \\
& - \frac{C_{1}}{\Lambda} (2\delta \zeta - \delta \xi) \, .
\label{gthetatheta}
\end{align}
\end{itemize}

\begin{itemize}
\item $(t,t)$ component of (\ref{fperturb}) 
\begin{align}
0=&\delta\varphi_{2}^{\prime \prime} -\tan t \delta \dot{\varphi_{2}} 
+ \frac{1}{\cos^{2}t}\delta \varphi_{2}  \nonumber \\
& - \frac{C_{1}}{2C^{2} \Lambda \cos^{2} t} (\delta \zeta - 2\delta \xi) \, .
 \label{ftt}
\end{align}
\item $(t,x)$ component of (\ref{fperturb}) 
\begin{align}
0=\delta \dot{\varphi}^{\prime}_{2} - \tan t \delta \varphi_{2}^{\prime} \, . \label{ftx}
\end{align}
\item $(x,x)$ component of (\ref{fperturb}) 
\begin{align}
0=&\delta \ddot{\varphi_{2}} -\tan t \delta \dot{\varphi_{2}} 
- \frac{1}{\cos^{2}t}  \delta \varphi_{2}  \nonumber \\
& + \frac{C_{1}}{2C^{2} \Lambda \cos^{2}t}  (\delta \zeta - 2\delta \xi) \, .
\label{fxx}
\end{align}
\item $(\theta, \theta)$, $(\phi, \phi)$ component of (\ref{fperturb}) 
\begin{align}
0=&2\delta \rho_{2}  
+\cos^{2}t \left(- \delta \ddot{\rho_{2}} + \delta \rho_{2}^{\prime \prime} 
+ \delta \ddot{\varphi_{2}} 
- \delta \varphi_{2}^{\prime \prime}\right)  \nonumber \\
& + \frac{C_{1}}{C^{2}\Lambda} (2\delta \zeta - \delta \xi) \, .
\label{fthetatheta}
\end{align}
\end{itemize}

\end{document}